%% file: TPC_Signal_MWPC.tex
\documentclass[a4paper,11pt]{article}
\pdfoutput=1 
\usepackage{jinstpub} 

\usepackage{hyperref}
\usepackage{lineno}
\usepackage{dcolumn}
\usepackage{bm}
\usepackage{amsmath}

\usepackage[utf8]{inputenc}
\usepackage{listings}
\usepackage{xcolor}
\usepackage{enumitem}
\usepackage{todonotes}
\usepackage{multirow}
\usepackage{relsize}
\usepackage{environ}
\usepackage{placeins}
\usepackage{float}
\usepackage{xspace}
\usepackage{needspace}

\newcommand{\dEdx}{d$E$/d$x$\xspace}
\newcommand{\Figref}[1]{Figure~\ref{#1}}
\newcommand{\figref}[1]{figure~\ref{#1}}
\newcommand{\NeCOtwo}{Ne-CO$_2$ (90-10)\xspace}
\newcommand{\NeCo}{Ne-CO$_2$\xspace}
\newcommand{\ArCOtwo}{Ar-CO$_2$ (90-10)\xspace}
\newcommand{\ArCo}{Ar-CO$_2$\xspace}
\newcommand{\imagefolder}{./figures}
\sloppypar
                           
\title{Signal shapes in multiwire proportional chamber-based TPCs}
\abstract{\input{abstract-eng}}
\collaboration{ALICE TPC Collaboration}
\input{./authors}

\arxivnumber{xxxxx} 

\begin{document}
\maketitle
\flushbottom

\section{\label{sec:intro}Introduction}

A charged particle traversing the gas in a TPC~\cite{Charpak:1973mug, Rolandi:1984pwe, Macdonald:1984sei, Alme:2010ke, ALICE:2000jwd} leaves a trace of ionization along its trajectory. The ionization electrons drift to the end plates, where gas amplification takes place. In conjunction with the recorded drift time of the electrons, the TPC provides a complete 3D image of the ionization deposited in the active detector volume.

The ALICE TPC~\cite{Alme:2010ke} is divided into two halves by a central drift electrode, and readout chambers are located at the end plates on both sides. Each end plate is subdivided into 18 azimuthal sectors. Each sector is subdivided into small inner readout chambers (IROCs) and larger outer readout chambers (OROCs).
During Runs 1 and 2 (until 2018) at the CERN Large Hadron Collider (LHC) and for the ALICE TPC, the amplification was provided by multiWire proportional chambers (MWPC)~\cite{Sauli:117989} with cathode pad readout. Inside the amplification region, defined by the cathode wires, the anode wires, and the pad plane (see figure~10 of ref.~\cite{Alme:2010ke}), avalanche signals are generated by the ionization electrons. A gating grid is located between the cathode wires and the drift region to prevent positive ions generated during the amplification process from entering the drift volume and causing field distortions.

The ions produced in the amplification process that are moving around the anode wires induce a current on the pads, which is picked up by the readout electronics, amplified further, and shaped with shaping time of about 190\,$ns$~\cite{ ALICE:2000jwd}. These ions drift away from the anode wires in different directions, at much lower drift velocities than the electrons, and are eventually collected at the pad plane, the cathode wires, or the gating grid~\cite{Rossegger:2010zz}. Their slow drift results in a very long signal tail with negative polarity.
This extended tail, along with the associated average baseline shift and fluctuations, leads to a significant degradation of the measured specific energy loss (\dEdx) and consequently of the particle identification (PID) performance of the TPC. The effect was reproduced at the digitization level in simulations based on the official ALICE simulation framework.  The effect was corrected during offline reconstruction of the data recorded with the TPC. We have included in this paper a discussion of the PID performance of the correction procedure based on simulated data, while more detailed information regarding collision data can be found in ref.~\cite{Arslandok:2022dyb}.

The first detailed study of the ion tail was performed using cosmic tracks and is described in ref.~\cite{Rossegger:2010zz}. In this paper we perform a more differential study of the ion-tail characteristics by analyzing ionization tracks generated by the ALICE TPC laser calibration system~\cite{Renault:2005tr}. Moreover, we extended the simulation studies to a full three-dimensional model. 

\section{\label{sec:data}Measurements with the ALICE TPC}

\begin{figure}[h]
  \centering
  \includegraphics[width=\textwidth]{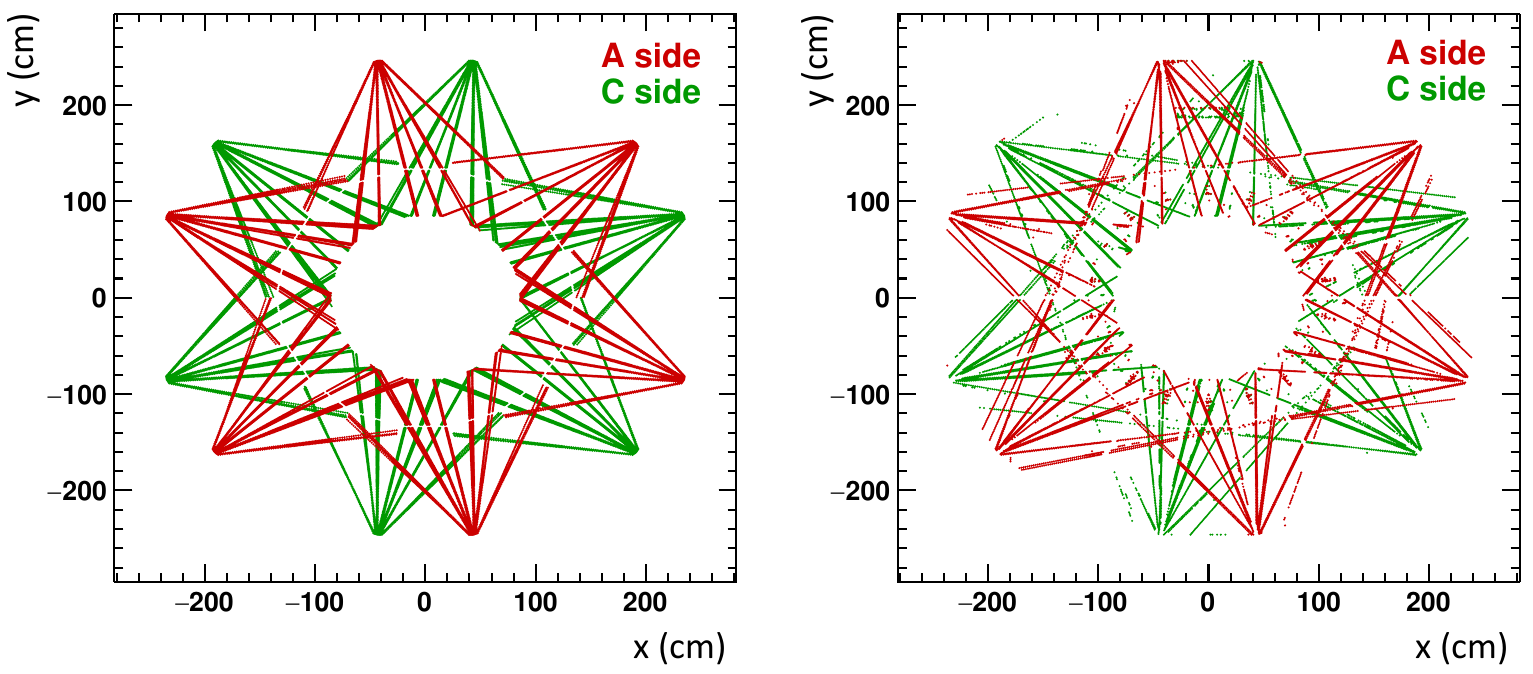}
  \caption{The ($x,y$) projection of the ideal laser positions (left) and the reconstructed laser tracks (right). The pattern repeats eight times along the entire length of the TPC, four times on each side (A and C) of the central electrode.}
  \label{fig:lasertracks}
\end{figure}

\begin{figure}[h]
  \centering
  \includegraphics[width=7.5cm]{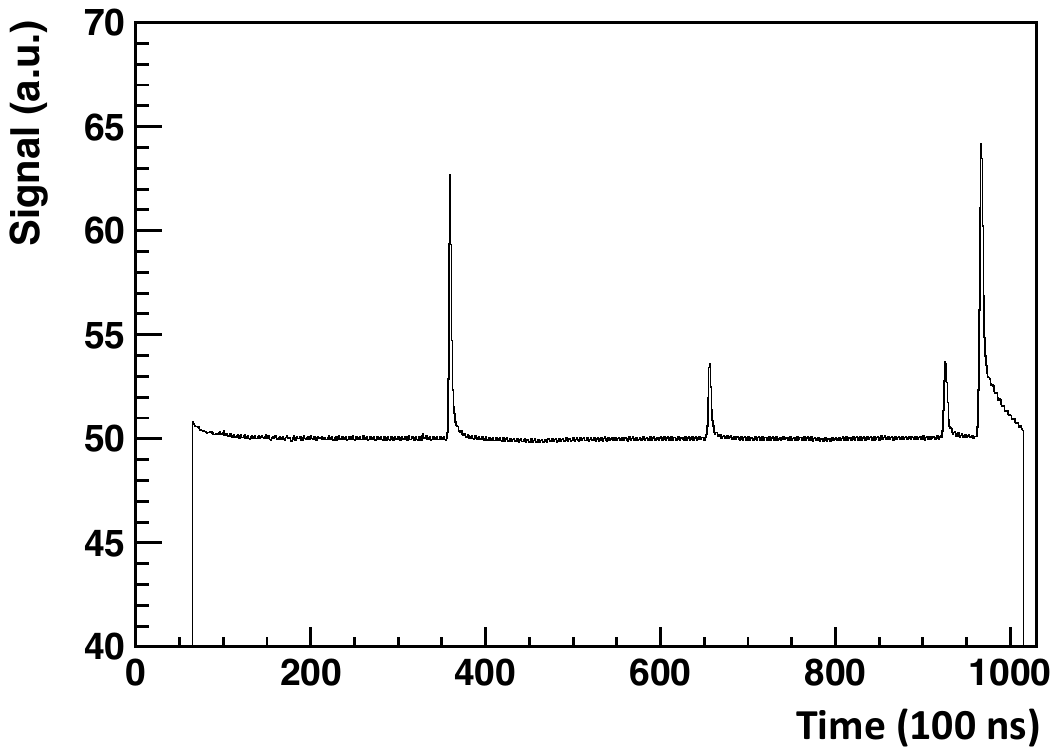}
  \caption{TPC laser signals on a single pad before pedestal subtraction and zero suppression over the entire drift time. The peaks at 36\,$\mu$s, 68\,$\mu$s, and 95\,$\mu$s correspond to the second, third, and fourth laser bundles, respectively, while the peak at 98\,$\mu$s corresponds to the central electrode signal originating from photoelectrons released from the central drift electrode. The first laser bundle closest to the pad plane is located at 13\,$\mu$s but no signal is evident on this pad in this event.}
  \label{fig:rawlaser_signals}
\end{figure}

The laser calibration system is used to generate 336 straight tracks parallel to the pad plane at known locations in the drift volume of the TPC (see figure~49 of ref.~\cite{Alme:2010ke}). In each half of the TPC, six broad laser beams illuminate 24 bundles of micro-mirrors positioned nearly equidistantly along the drift-time direction. Within each bundle, seven narrow beams are reflected into the TPC. The projections of the laser tracks on the pad plane ($x,y$ coordinates) and on the time axis ($z$ coordinate)  are shown in \figref{fig:lasertracks} and \figref{fig:rawlaser_signals}, respectively. 
 
The laser system uses a pulsed UV laser beam with a wavelength of 266\,nm from an Nd:YAG laser. It is designed to provide 100\,mJ per pulse of 5\,ns duration at a repetition rate of 10\,Hz. The data analysed in this publication were taken in dedicated standalone calibration runs where the event readout was triggered by the laser system.
This mode of operation provides the flexibility to read all data without suppression of the baseline for the individual electronics channels, and to tune several external parameters, such as the anode wire voltage, the number of events, and the laser intensity.

\begin{figure}[h]
  \centering
    \includegraphics[width=\linewidth]{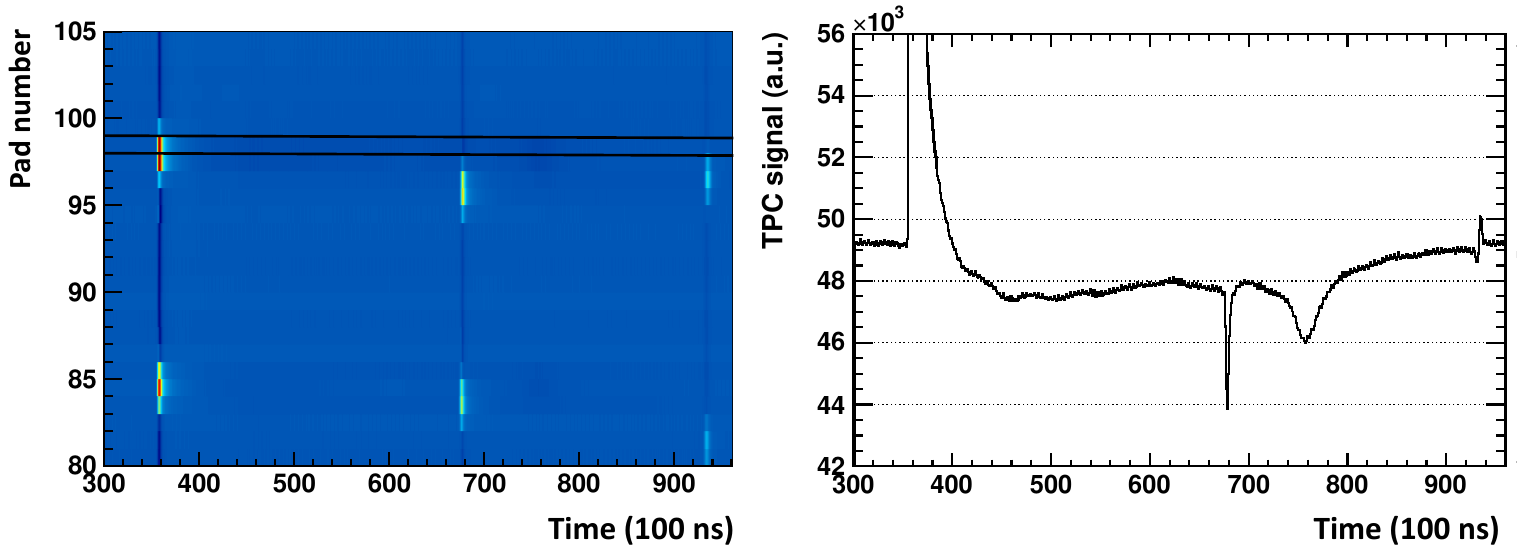}
  \caption{\textit{Left:} Charge clusters from laser tracks in the pad-time plane. \textit{Right:} signal shape as a function of time on a single pad indicated by the two solid black lines in the left panel. The signal is summed over 1000 laser events. The sharp dip at around $68$\,$\mu$s corresponds to the common-mode signal which is explained in ref.~\cite{Arslandok:2022dyb}.}
  \label{fig:SigComp}
\end{figure}

\begin{figure}[h]
  \centering
    \includegraphics[width=8cm]{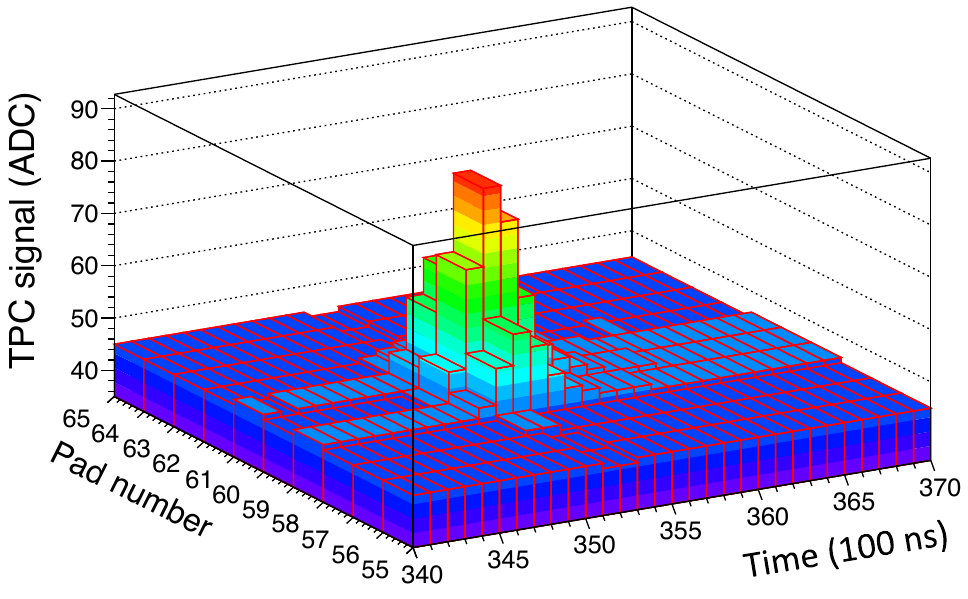}
  \caption{A laser cluster before pedestal subtraction and zero suppression.}
  \label{fig:LaserSig}
\end{figure}

There are several challenges in the signal shape analysis. Since the characteristic ion tail spreads over tens of microseconds, and the magnitude of the undershoot is comparable to the electronics noise, a data set with low track density is required such that the signal tail is not distorted. In addition, a high-statistics sample is needed to ensure good resolution of the tail shape. For these reasons cosmic tracks were used in the past~\cite{Rossegger:2010zz}. Laser data were not initially considered due to the repetitive structure of the laser tracks along the time axis and the shorter distances between the laser layers ($\sim 30$\,$\mu$s) compared to a typical signal tail ($\sim 40$\,$\mu$s for inner and $\sim 80$\,$\mu$s for outer readout chambers). However, the finite resolution in the angle of the laser mirrors in the ($x,y$) plane results in pad regions where at least one of the four laser signals is not recorded. These regions could then be used to inspect signals where the long ion tail was undistorted by subsequent large  positive laser signals. The example in the right panel of \figref{fig:SigComp} shows a signal at 38\,$\mu$s whose tail does not overlap with the subsequent laser pulse at 68\,$\mu$s, but is instead on the adjacent pad due to the above mentioned slight misalignment of the laser track positions. The six distinct hotspots on the left plot correspond to laser clusters originating from different laser tracks. Here, a cluster is defined as an accumulation of charge signals detected within a search window of five bins each in pad and time direction, as illustrated in \figref{fig:LaserSig}.

\begin{figure}[h]
  \centering
  \includegraphics[width=\linewidth]{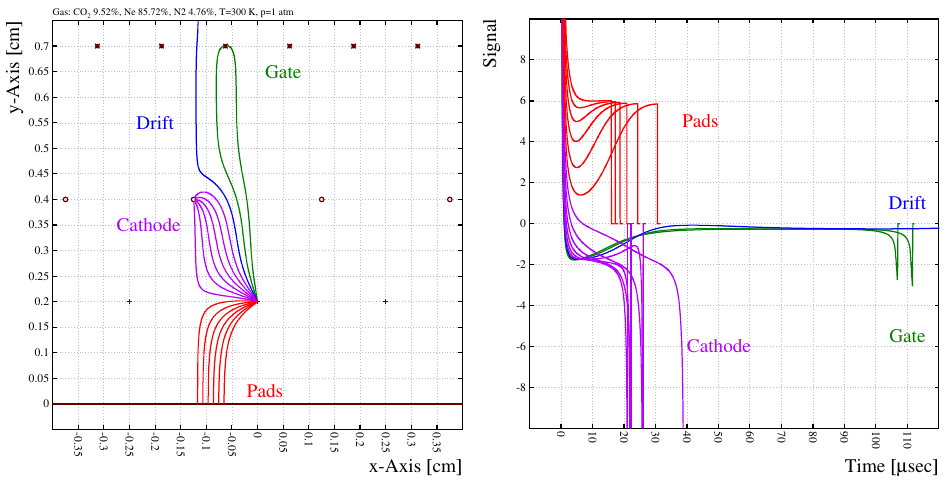}
  \caption{\textit{Left:} the drift curves of ions originating from the avalanche close to the anode wire. Ions collected on the pad plane, cathode wire, and gating grid are shown with red, purple, and green curves, respectively, while ions escaping back into the drift volume are shown in blue. \textit{Right:} corresponding ion signals induced on the cathode pads~\cite{Rossegger:2010zz}.}
  \label{fig:steffGarfield}
\end{figure}

Signal generation is well understood and implemented into the Garfield simulation package, which allows accurate modeling of the signal response, including primary ionization, electron drift, diffusion, and signal induction on different electrodes~\cite{garfieldref, Rossegger:2010zz}. The simulated drift paths and the corresponding signal shapes for individual ions produced in the vicinity of the anode wire are shown in \figref{fig:steffGarfield}. Ions moving towards the pad plane add positively to the signal tail, whereas ions drifting away contribute negatively. Thus, the slight increase of the signal between 45 and 62\,$\mu$s in the right panel of \figref{fig:SigComp} results from ions collected at the pad plane, and the dip at around 76\,$\mu$s is caused by ions collected on the cathode wires. 

The mobility of the electrons is about three orders of magnitude larger than the mobility of the ions, therefore electrons are essentially instantaneously removed from the amplification region. As a consequence, the signal induced on the cathode pads is characterized by a fast rise due to the ions generated at the high field in the vicinity of the anode wire and by a long tail due to the slow drift in the low-field region. While the maximum amplitude of the ion tail is about 1\% of the fast-signal pulse height, its extended duration means the ion-tail integral is comparable to that of the initial positive signal. Consequently, subsequent signals on the same readout pad that are close in time are strongly affected by signal overlap, resulting in a significant reduction of the signal quality. In particular, in high-multiplicity scenarios some fraction of the pad signal may fall below the zero suppression threshold and may be lost (see for example figure~6 of ref.~\cite{Arslandok:2022dyb}). Hence, the study of the signal shape dependencies played a crucial role in improving the calibration of energy-loss measurements in the TPC, as elaborated in ref.~\cite{Arslandok:2022dyb}.

\section{\label{sec:itcm}Characterization of the ion tail}

\subsection{\label{sec:iontalshape}Signal shape dependencies}

\begin{figure}[h]
  \centering
  \includegraphics[width=9cm]{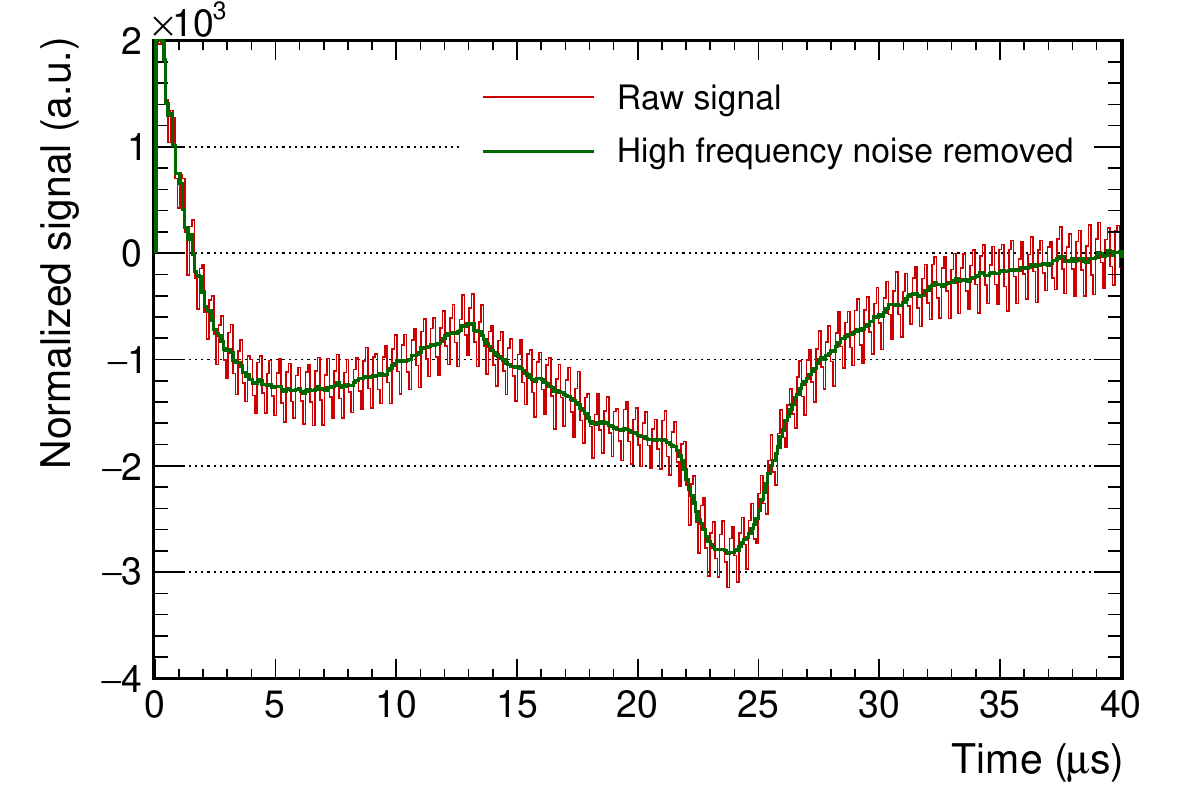}
  \caption{(Color Online) The ion-tail signal of a central pad for the IROC geometry before (red curve) and after (green curve) the removal of the high-frequency noise.}
  \label{fig:FFTsignal}
\end{figure}

To investigate the signal shapes thoroughly, we first performed a scan of approximately 0.5 million pads in the TPC to identify all signals with undistorted tails. We then improved the signal-to-noise ratio by doubling the number of events per run to 2000, compared to the default 1000 for a stand-alone laser run. In addition, we increased the laser intensity to obtain larger pulse heights, achieving signal amplitudes of about 800\,ADC at a noise level of about 1\,ADC. Furthermore, we identified an additional high-frequency electronic noise which we  filtered using the Fast Fourier Transform feature within the ROOT framework~\cite{Brun:1997pa}. The result after elimination of the high-frequency noise is depicted in \figref{fig:FFTsignal}. 

There are several aspects that determine the shape of the ion tail: the gas mixture, the electrode geometry, the field configuration, and the position of the given readout pad with respect to the center-of-gravity of the charge cluster. 

\begin{figure}[h]
  \centering
    \includegraphics[width=9cm]{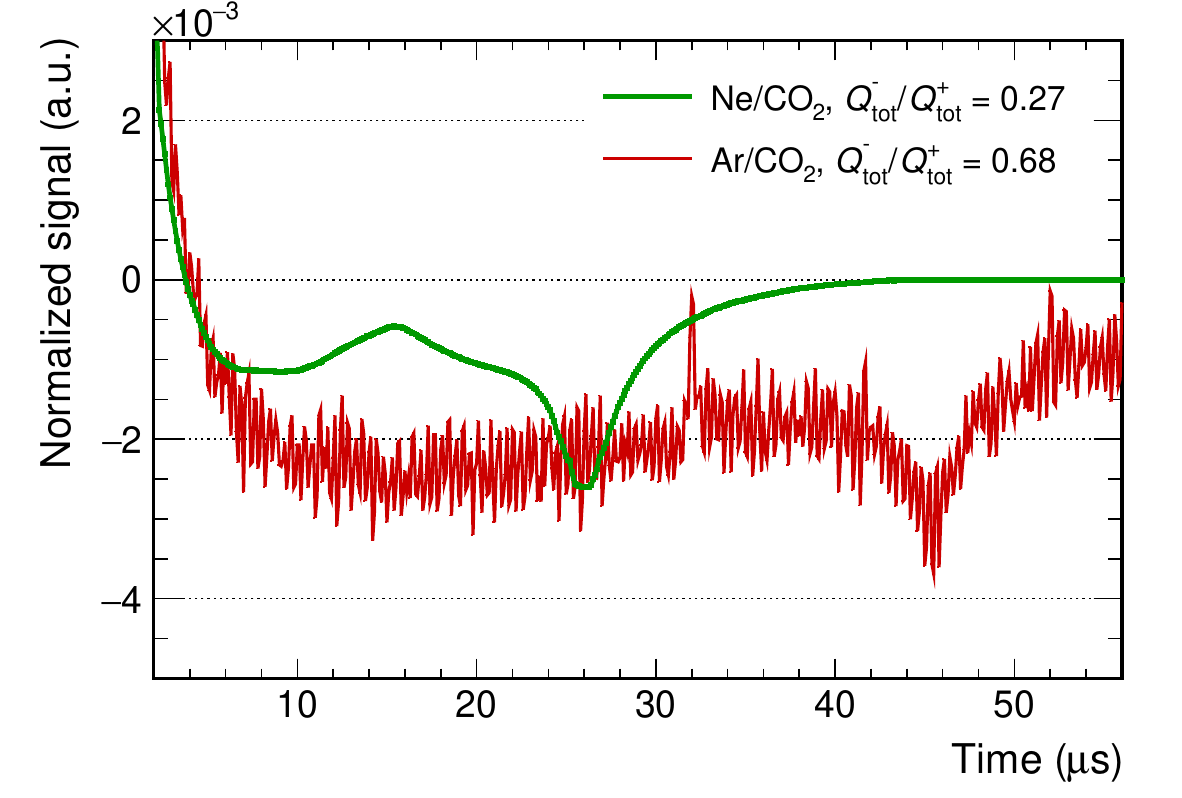}
  \caption{(Color Online) Normalized pad signals of the central pads for the IROC geometry with a zoom into the y-axis for \ArCOtwo (red) and \NeCOtwo (green) gas mixtures. The \NeCo data have a factor of 20 larger statistics than the \ArCo data, where no removal of high-frequency noise was performed.}
  \label{fig:gasDep}
\end{figure}

We first investigated the dependence on the gas mixture. \Figref{fig:gasDep} shows the ion-tail signal shapes in the \ArCOtwo and \NeCOtwo gas mixtures, where significant differences are observed. The primary ionization in \ArCo is about two times higher than in Ne-CO$_2$~\cite{Workman:2022ynf}. Therefore, the gas gain for \ArCo had to be kept correspondingly lower to keep the amplitudes of the fast-signal pulses  the same for the two gas mixtures. This resulted in a significant change in the avalanche spread around the anode wires, which in turn altered the number of ions directed towards the drift volume as opposed to those collected at the pad plane. This is reflected in a difference between the ratios of the integral of the negative ion tail (``$Q_{\rm tot}^-$") and the integral of the positive signal (``$Q_{\rm tot}^+$"). For the \ArCo the $Q_{\rm tot}^-/Q_{\rm tot}^+$ ratio was 0.68, while for the \NeCo mixture it is much lower at 0.27. In addition, the shift of the second minimum, from $\sim$\,26\,$\mu$s  for \NeCo to $\sim$\,45\,$\mu$s  for \ArCo, can be explained by the different ion arrival times at the cathode wires due to the different ion mobilities, which are $\sim$\,3.13\,$\mathrm{cm}^2/\mathrm{Vs}$ and $\sim$\,1.61\,$\mathrm{cm}^2/\mathrm{Vs}$, respectively (see ref.~\cite{Kalkan:2015vkb}), which uses the current data for the ion mobility calculations). These values are consistent with the ion mobility measurements reported elsewhere~\cite{Deisting:2018vtx}. For the remainder of the paper, we analyzed signal shapes generated in the \NeCo gas mixture due to their superior statistical quality.

\begin{figure}[h]
  \centering
    \includegraphics[width=9cm]{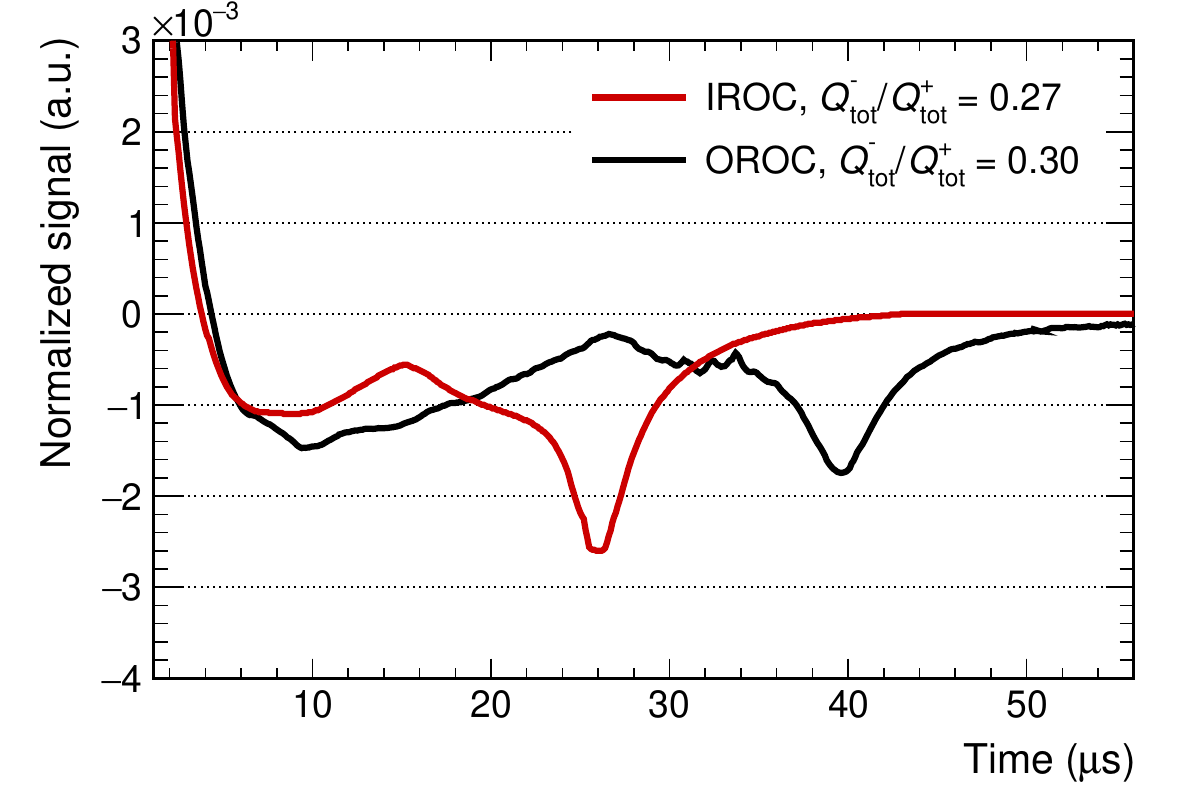}
  \caption{(Color Online) Normalized pad signals of the central pads with a zoom into the $y$-axis for IROC and OROC geometries, for the anode voltage settings of 1270\,V and 1490\,V respectively.}
  \label{fig:rocDep}
\end{figure}

The ALICE TPC employs different wire geometries for IROCs and OROCs, where the difference lies in the gap sizes between the cathode wires, anode wires, and pad plane (see figure~10 of ref.~\cite{Alme:2010ke}). In \figref{fig:rocDep}, normalized signals are shown for IROC and OROC geometries where the $Q_{\rm tot}^-/Q_{\rm tot}^+$ ratio and consequently the gain and the avalanche spread around the anode wire are similar. Due to the larger distances between the anode-wire plane and the cathode-wire plane as well as between the anode-wire plane and the pad plane, which lead to larger ion drift times, the time positions of the local maximum and of the second minimum shift to larger values in the case of the OROC geometry. 

\begin{figure}[h]
  \centering
  \includegraphics[width=1\linewidth]{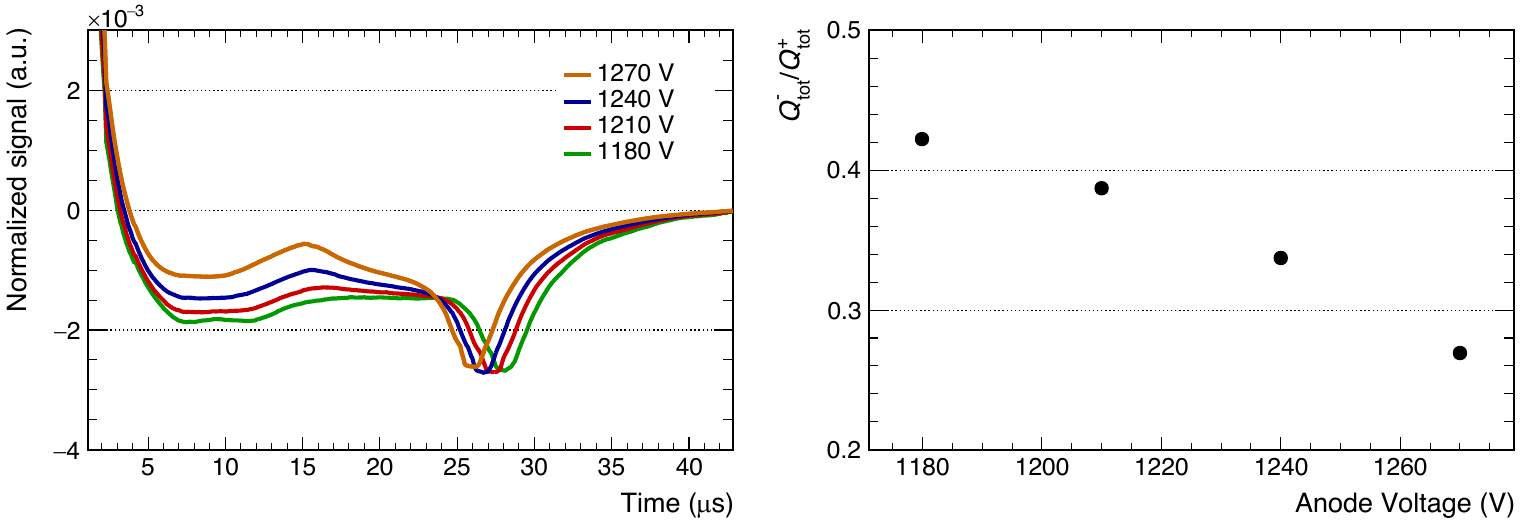}
  \caption{(Color Online) Normalized pad signals of the central pads for the IROC geometry with a zoom into the $y$-axis for different anode voltage settings. \textit{Left:} ion-tail shape. \textit{Right:} the $Q_{\rm tot}^-/Q_{\rm tot}^+$ ratio of the signal as a function of anode voltage.}
  \label{fig:voltageDep}
\end{figure}

Since the ion drift velocity increases with higher fields, the anode wire voltage also has an effect on the position of the local minima, as shown in the left panel of~\figref{fig:voltageDep}. However, the more dominant effect reveals itself in a change of the avalanche spread around the anode wire. An increase in the anode wire voltage not only leads to higher gains but also an earlier start of the avalanche, and hence an enlarged angular spread of the avalanche around the anode wire. This increases the number of ions arriving at the pad plane, as shown as red solid lines in~\figref{fig:steffGarfield}. As a consequence, the second local maximum at about 15\,$\mu$s is enhanced and the $Q_{\rm tot}^-/Q_{\rm tot}^+$ ratios are reduced with increasing anode voltage, as shown in~\figref{fig:voltageDep}.

\begin{figure}[h]
  \centering
  \includegraphics[width=1\linewidth]{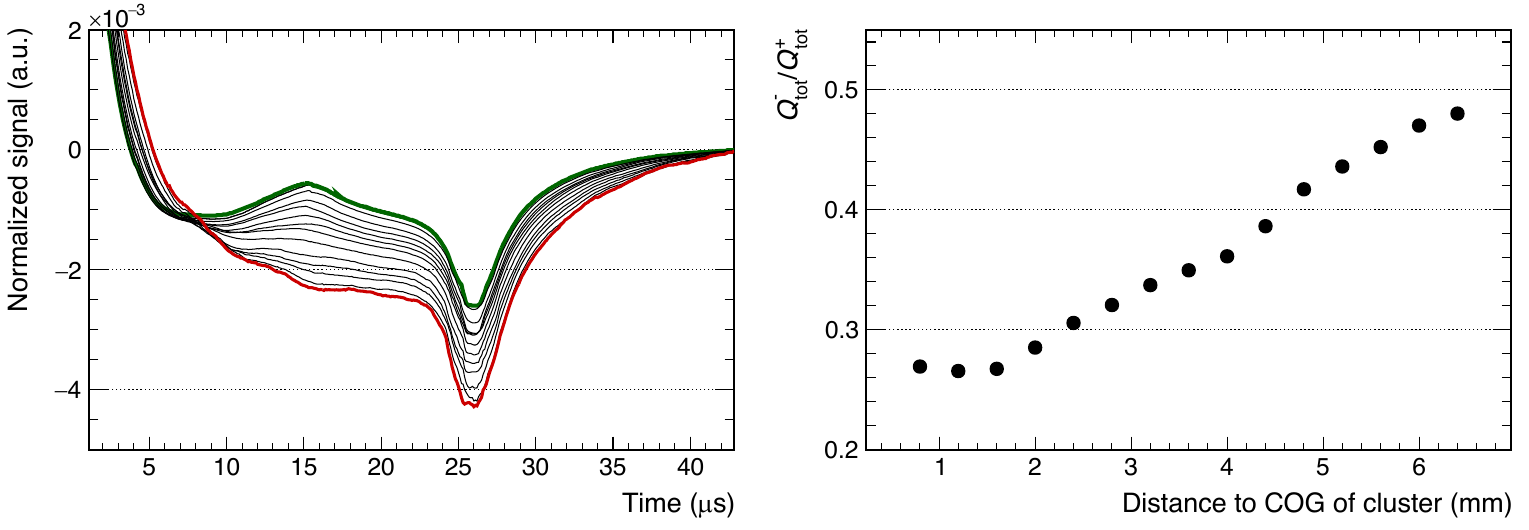}
  \caption{Normalized pad signals for the IROC geometry with a zoom into the y-axis measured at different distances to the center-of-gravity (dcog) of the cluster (anode voltage = 1270\,V). \textit{Left:} Ion-tail shape where solid green and red curves indicate the most central (dcog\,$=0$\,$mm$) and peripheral (dcog\,$=16$\,$mm$) pad signals, respectively. \textit{Right:} the $Q_{\rm tot}^-/Q_{\rm tot}^+$ ratio of the signal as a function of distance to the center-of-gravity of the cluster.}
  \label{fig:cogDep}
\end{figure}

While the anode voltage, wire geometry, and gas mixture remain fixed parameters during collision-data taking, there is another strong dependence that must be taken into account: the distance of a given pad  from the center-of-gravity (COG) of the cluster, which varies for every measured signal. Each pad within a given cluster receives a different fraction of ions produced during amplification, as well as a different magnitude of induced mirror charge due to the avalanche spread around the anode wires. This leads to differences in the shape of the ion tail of each pad in the cluster, as shown in~\figref{fig:cogDep}, where a dramatic change in the $Q_{\rm tot}^-/Q_{\rm tot}^+$ ratio is observed when moving from the most central pads to the more peripheral ones.

\subsection{\label{sec:iontailsim}Simulations}

\begin{figure}[h]
  \centering
    \includegraphics[width=0.65\linewidth]{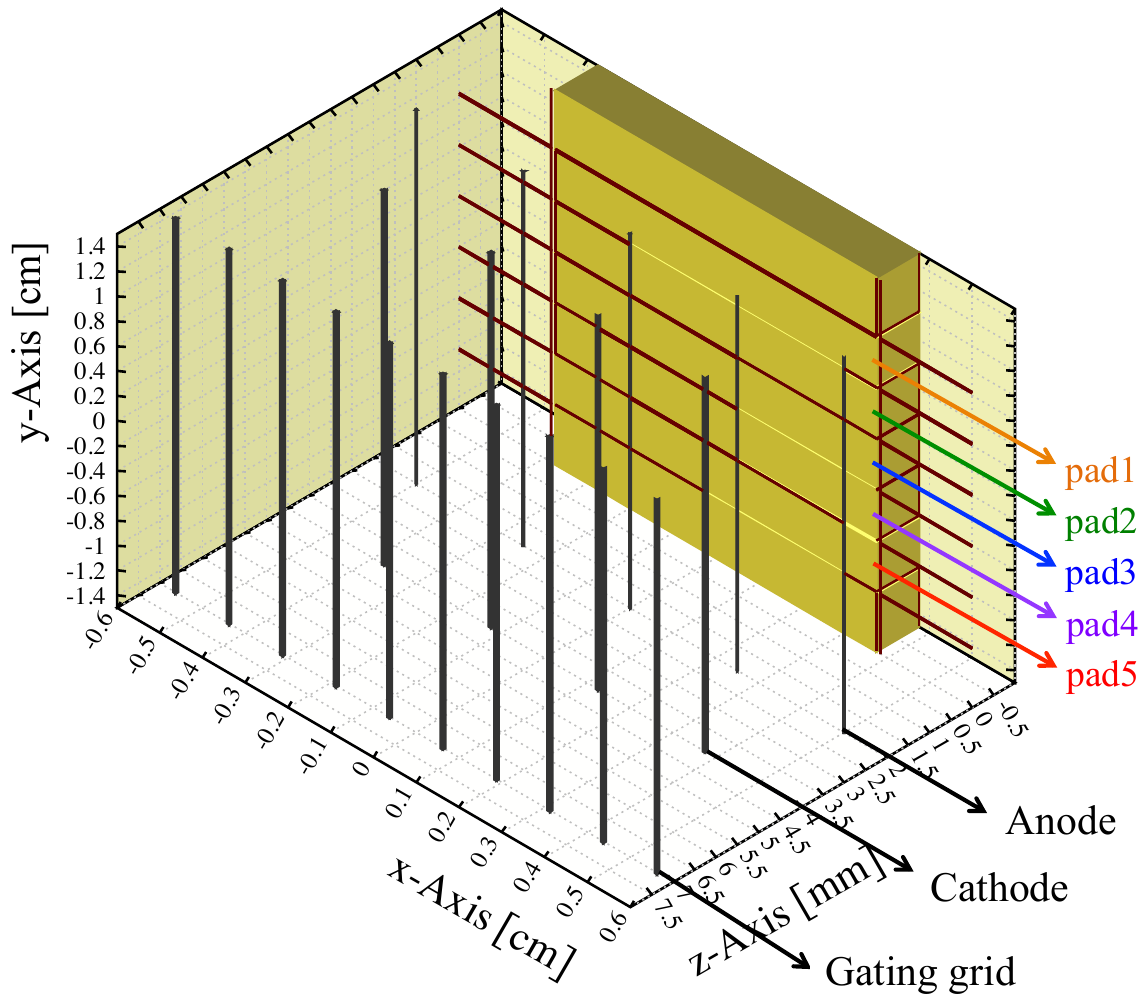}
  \caption{Three-dimensional setup of the ALICE TPC IROC wire and pad geometry used in the Garfield simulations. The black lines on the pad plane serve to mark the boundaries between individual pads.}
  \label{fig:GarfieldSetup}
\end{figure}

In its original implementation, Garfield was designed to model two-dimensional chamber geometries made of wires and planes where the exact fields are known. However, for three-dimensional configurations, the dielectric media and complex electrode shapes are difficult to handle with analytic techniques. To cope with this problem, Garfield is interfaced with the neBEM (A nearly exact Boundary Element Method)~\cite{nebemref} program, which provides the field maps as the basis for the calculations of Garfield. The three-dimensional wire and pad geometry of the ALICE TPC IROC created using the neBEM program is shown in \figref{fig:GarfieldSetup}. In addition to the pads that are analysed in the simulation, two pads were added on each side to avoid boundary effects.

\begin{figure}[h]
  \centering
  \includegraphics[width=\linewidth]{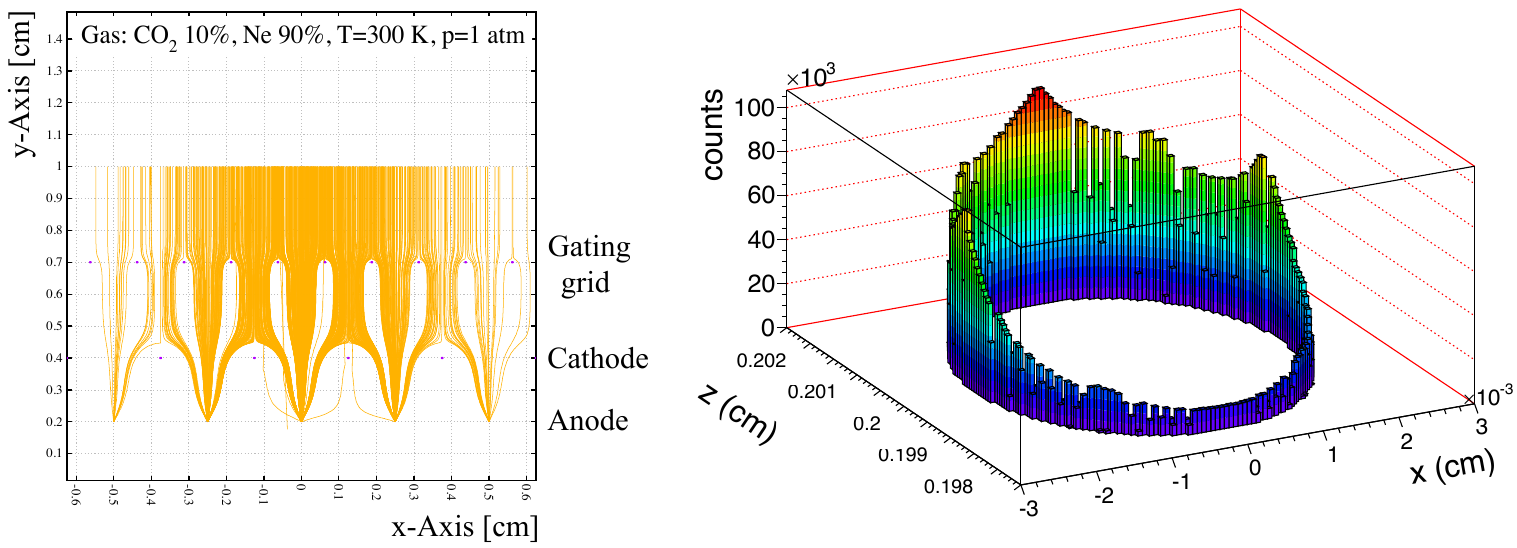}
  \caption{\textit{Left:} drift paths of electrons in the amplification region (without diffusion), where the arriving primary electrons are assumed to be distributed according to a Gaussian. \textit{Right:} the resulting ion distribution around the anode wire at the start of drift.}
  \label{fig:IonTail_elDrift}
\end{figure}

The drift path of the ions, and hence the ion signal induced on the pads, depends on their distribution around the anode wire. To this end, an electron cloud with Gaussian density profile is generated outside the amplification region. The electrons are then drifting through the wire grids to the anode wires, neglecting diffusion, as shown on the left panel of \figref{fig:IonTail_elDrift}. The arrival points of the electrons at the anode wire are used to estimate the ion distribution around the wire, see \figref{fig:IonTail_elDrift}, right panel. 

\begin{figure}[h]
  \centering
  \includegraphics[width=0.78\linewidth]{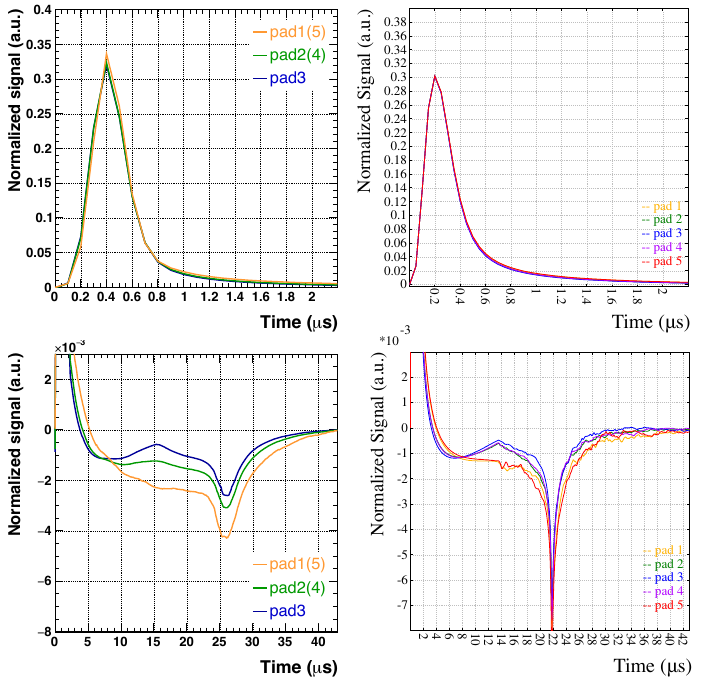}
  \caption{Comparison of the measured and simulated signal shapes for the IROC geometry in five pads of a cluster where pad number 3 is the center of the cluster. \textit{Upper panel:} pulse shape. \textit{Lower panel:} ion tail. \textit{Left:} data. \textit{Right:} Garfield simulation.}
  \label{fig:CompGarfieldSetup}
\end{figure}

Reasonable agreement between the data and Garfield simulations is achieved, as shown in \figref{fig:CompGarfieldSetup}. In particular, the pad position dependence is reproduced.  However, the simulated signal shapes exhibit a sharp negative spike. The smearing of these spikes in the real data can be attributed to the fluctuations in wire positions, which are assumed to be Gaussian and are in the range of $\sigma_{\rm geom}\sim50$\,$\mu$m~\cite{ALICE:2000jwd, Rossegger:2010zz}. 

\begin{figure}[h]
  \centering
    \includegraphics[width=9cm]{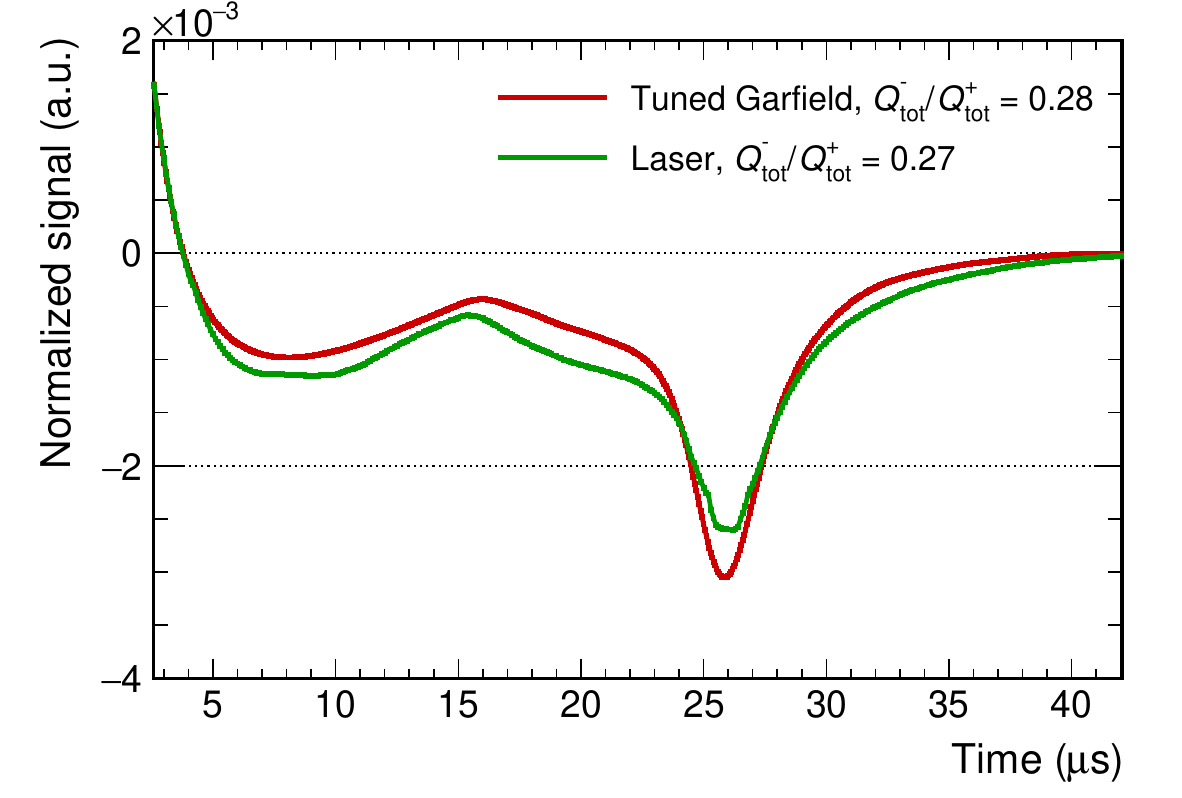}
  \caption{Comparison of the normalized pad signal for the IROC geometry to the Garfield simulations in a \NeCOtwo gas mixture for the most central pad of a cluster. The simulation result is scaled to match the measured ion drift velocity and also smeared, taking the fluctuations of the wire positions into account.}
  \label{fig:smearedGarf}
\end{figure}

Furthermore, there is a striking difference in the timing of the negative spike: it appears at about 22\,$\mu$s in the simulations and 26\,$\mu$s in the data. This is mainly due to the lack of information in the literature about the identity of the drifting ions in a \NeCOtwo gas mixture. Therefore, in our current simulation we have opted for a simplified approach, considering only the presence of noble gas ions (Ne$^{+}$) and neglecting the influence of the quencher (CO$_2$) as a reasonable approximation. In ref.~\cite{Kalkan:2015vkb}, it is shown that CO$_2$ cluster mobilities, rather than that of Ne$^{+}$, can describe the observed data. Eventually, by accounting for variations in wire positions and ion distributions around the anode wires, as well as using the measured ion mobility, we have achieved an excellent agreement between simulation and data, as shown in \figref{fig:smearedGarf}. Note that this analysis covers only one type of topology (i.e., cluster shape) where the track is parallel to the pad plane. Since high-precision, undistorted signals are very difficult to obtain using collision data, one can use such a data-driven approach to cover all different track topologies by relying on the three-dimensional simulations, mainly tuning the ion distributions around the anode wires.

\section{Impact of the ion-tail correction on the PID performance}

The PID performance of the ALICE TPC is deteriorated as a consequence of the ion tail. A quantitative study is performed employing the ALICE TPC simulation framework where the observed ion tail shapes are implemented into the simulation. Collisions of Pb ions are generated using the HIJING event generator~\cite{Gyulassy:1994ew} and propagated through the detector setup using GEANT4~\cite{GEANT4:2002zbu} transport software. During the reconstruction of the simulated data, the total charge of each cluster within a given track was corrected by accounting for the baseline shift which was simulated with realistic ion tail shapes. 

The baseline shift estimate is based on the normalized pad signals obtained from the laser data mentioned above. The corrected pad signals for a given cluster are approximated with the following vector operation:
\begin{equation}
 \mathbf{S_{\rm corr}} = \mathbf{S_{\rm in}}  +\alpha_{\rm IT} \cdot Q_{\rm tot}\cdot  \mathbf{S_{\rm norm}},
\end{equation} 
where $\mathbf{S_{\rm corr}}$ and $\mathbf{S_{\rm in}}$ are the corrected and input pad signals, respectively, $Q_{\rm tot}$ is the integral of the positive part of the input pad signal, $\mathbf{S_{\rm norm}}$ is the normalized pad signal for the given relative pad position to the center-of-gravity of cluster, and $\alpha_{\rm IT}$ is a fudge factor accounting for the missing charge and clusters due to zero suppression. After cluster finding the coordinates of the center-of-gravity, $Q_{\rm tot}$ and $Q_{\rm max}$ of each cluster are available but detailed signal shape information for individual pads is not retained. To address this, the shape of a specific cluster is estimated using the "pad response function" characterized by a Gaussian shape in both time and pad direction for simplicity. This approach allows a straightforward calculation of the relative position of each pad along with its associated $Q_{\rm tot}$. Eventually, the estimated baseline shift is incorporated into each cluster as missing charge. Applying such a correction to each individual cluster cannot restore signals that fall below the zero suppression threshold. However, it can restore the lost amplitude for the detected signals. On the other hand, it should be noted that certain imperfections are unavoidable. For example, the normalized signal shapes are generated by laser tracks that are parallel to the pad plane. Therefore, variations in the cluster shape due to track angles cannot be accounted for. 

\begin{figure}[h]
  \centering
    \includegraphics[width=9cm]{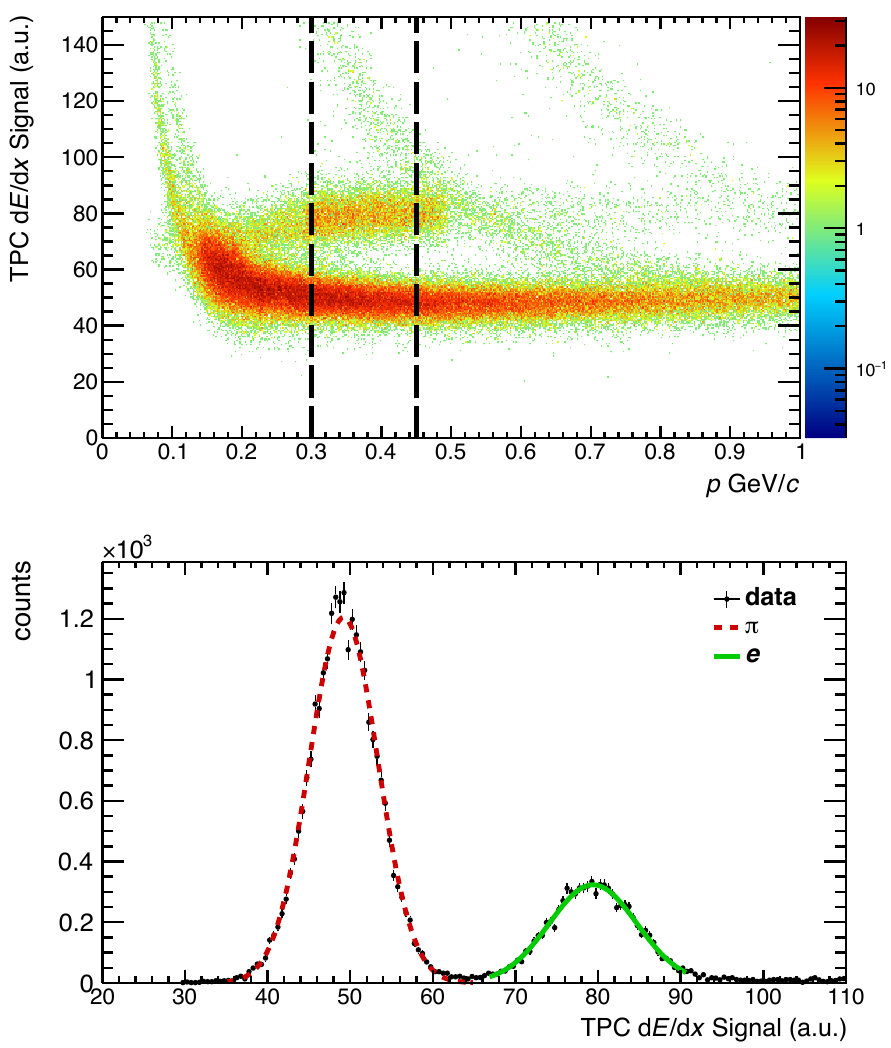}
  \caption{\textit{Upper panel:} the \dEdx spectrum of simulated central Pb--Pb events obtained via HIJING simulation with GEANT4 implementation of the TPC detector setup, where electrons are enhanced by embedding $\gamma$ conversions in order to obtain sufficient statistics for the fit. Dashed curves depict the momentum range where the separation power calculation is performed. \textit{Lower panel:} the projection of the marked area of the upper plot on the \dEdx axis.The Gaussian fits for pions and electrons are shown by red dashed and green solid lines, respectively. The resulting separation power is $6.4\pm0.04$.}
  \label{fig:sep_power}
\end{figure}

As a measure of the PID performance we use the so-called separation power $S_{e,\pi}$ between electrons and minimum-ionizing pions:
\begin{equation}
  S_{e,\pi} = \frac{|\mu_{\pi}-\mu_{e}|}{\frac{1}{2} ( \sigma_{\mathrm{d}E/\mathrm{d}x,e} + \sigma_{\mathrm{d}E/\mathrm{d}x,\pi} )}, \\
\end{equation}
where $\mu_{i}$ and $\sigma_{\mathrm{d}E/\mathrm{d}x,i}$ are the average \dEdx signal and \dEdx resolution of particle $i$ (electron $e$ and pion $\pi$) within a given momentum range, respectively. The separation power inherently incorporates information about the \dEdx resolution and serves as the main parameter for evaluating the PID performance of the TPC. Figure \ref{fig:sep_power} provides a visual illustration of how separation power is calculated, using a small statistics test of a HIJING simulation implemented with the GEANT4 TPC detector setup for central Pb--Pb collisions. 

\begin{figure}[h]
  \centering
    \includegraphics[width=\linewidth]{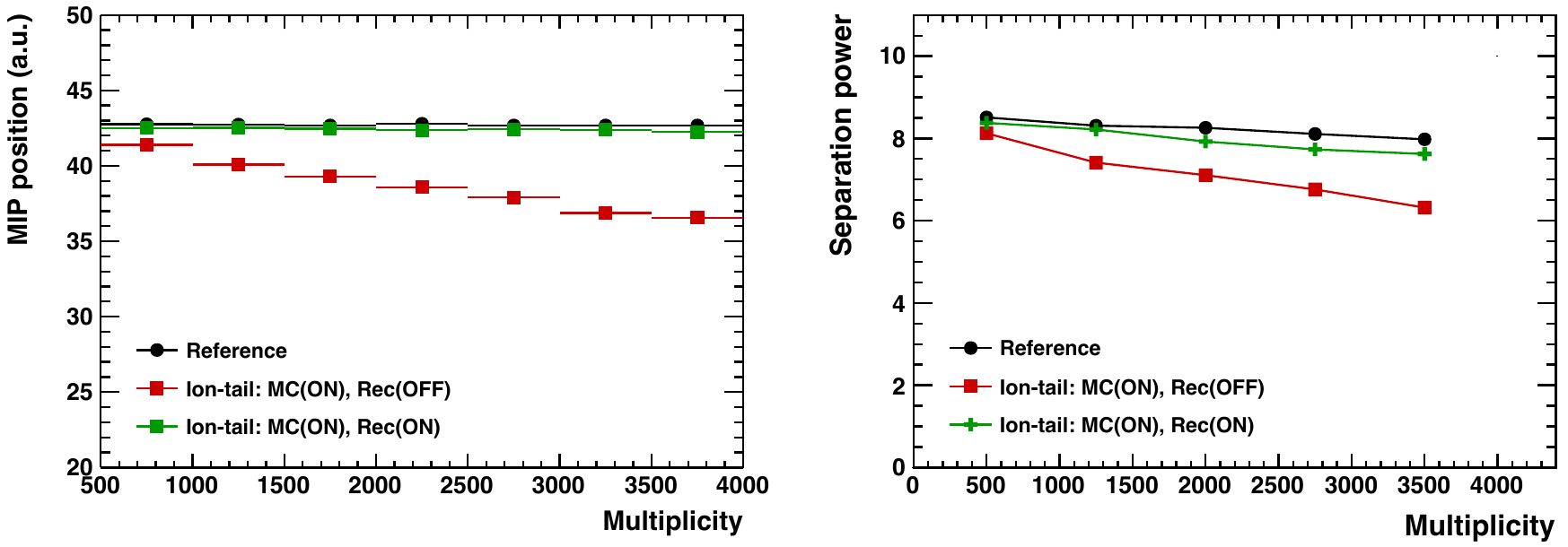}
  \caption{\textit{Left:} mean \dEdx position of minimum-ionizing pions and separation power obtained from a HIJING simulation including the ion-tail effect. The red and green symbols show the achieved performance before and after offline correction as a function of the event multiplicity, respectively. The black symbols show the performance without any tail.}
  \label{fig:mip_pos}
\end{figure}

The electron-pion separation power and the mean \dEdx position of the minimum ionizing pions as a function of multiplicity are shown in \figref{fig:mip_pos}. The number of primary particles in the collision is used as the multiplicity estimator. The reference simulation data without ion tail show a slight decrease with increasing multiplicity, which is due to the merging of clusters at high detector occupancies. On the other hand, the ion-tail effect leads to a decrease of approximately 13\% in the MIP position and about 22\% in separation power, as shown in \figref{fig:mip_pos}. These effects are almost completely recovered by an offline procedure where the known ion-tail shapes are used to correct each signal according to the properties of the preceding signals. The remaining imperfections can be attributed to signals that are lost during the zero suppression. These signals can not be recuperated with the offline correction, because the offline correction is applied on the cluster level after zero suppression.

\section{\label{sec:conclusion}Conclusions}

A detailed study of signal shapes in the MWPCs of the ALICE TPC was carried out using data recorded with the laser calibration system. It was found that the signals have a characteristic long negative ion tail after the main positive signal. The tail strongly depends on the gas mixture, the electrode geometry, the field configuration, and the position of a particular pad signal with respect to the center-of-gravity of the cluster. The characteristic signal shape was verified for the first time with three-dimensional Garfield simulations. Ion tails of this nature can be found in all MWPC-based TPCs and it is important to consider them when operating in high multiplicity environments to achieve good tracking and PID performance, as demonstrated through a comprehensive detector simulation that incorporates these ion tails. The results obtained have not only extended the understanding of the signal shape dependencies, but also significantly improved the performance of the offline correction procedure for Pb--Pb collision events, as reported in ref.~\cite{Arslandok:2022dyb}. 

\section*{Acknowledgements}
The ALICE TPC Collaboration acknowledges the support of the following funding agencies:
Funda\c{c}\~ao de Amparo \`a Pesquisa do Estado de S\~ao Paulo (FAPESP), Brasil;
Ministry of Science and Education, Croatia;
The Danish Council for Independent Research | Natural Sciences, the Carlsberg Foundation and Danish National Research Foundation (DNRF), Denmark;
Helsinki Institute of Physics (HIP) and Academy of Finland, Finland;
Bundesministerium f\"{u}r Bildung, Wissenschaft, Forschung und Technologie (BMBF), 
GSI Helmholtzzentrum f\"{u}r Schwerionenforschung GmbH, 
DFG Cluster of Excellence "Origin and Structure of the Universe", 
The Helmholtz International Center for FAIR (HIC for FAIR)
and the ExtreMe Matter Institute EMMI at the GSI Helmholtzzentrum f\"{u}r Schwerionenforschung, Germany;
National Research, Development and Innovation Office, Hungary;
Nagasaki Institute of Applied Science (IIST)
and the University of Tokyo, Japan;
Fondo de Cooperaci\'{o}n Internacional en Ciencia y Technolog\'{i} a (FONCICYT), Mexico;
The Research Council of Norway, Norway; Ministry of Science and Higher Education and National Science Centre, Poland;
Ministry of Education and Scientific Research, Institute of Atomic Physics and Ministry of Research and Innovation, and Institute of Atomic Physics, Romania;
Ministry of Education, Science, Research and Sport of the Slovak Republic, Slovakia;
Swedish Research Council (VR), Sweden;
United States Department of Energy, Office of Nuclear Physics (DOE NP), United States of America. 










\bibliographystyle{utphys}   
\bibliography{./biblio.bib}

\end{document}

%% file: abstract-eng.tex

A large-volume Time Projection Chamber (TPC) is the main tracking and particle identification (PID) detector of the ALICE experiment at the CERN LHC. PID in the TPC is performed via specific energy-loss measurements (\dEdx), which are derived from the average pulse-height distribution of ionization generated by charged-particle tracks traversing the TPC volume. During Runs\,1 and 2, until 2018, the gas amplification stage was based on multiwire proportional chambers (MWPC). Signals from the MWPC show characteristic long negative tails after an initial positive peak due to the long ion drift times in the MWPC amplification region. This so-called ion tail can lead to a significant amplitude loss in subsequently measured signals, especially in the high-multiplicity environment of high-energy Pb--Pb collisions, which results in a degradation of the \dEdx resolution. A detailed study of the signal shapes measured with the ALICE TPC with the \NeCOtwo and \ArCOtwo gas mixtures is presented, and the results are compared with three-dimensional Garfield simulations. The impact of the ion tail on the PID performance is studied employing the ALICE simulation framework and the feasibility of an offline correction procedure to account for the ion tail is demonstrated.

%% file: authors.tex
\affiliation[1]{Comenius University Bratislava, Faculty of Mathematics, Physics and Informatics, Bratislava, Slovak Republic}    
\affiliation[2]{Department of Physics and Technology, University of Bergen, Bergen, Norway}          
\affiliation[3]{European Organization for Nuclear Research (CERN), Geneva, Switzerland}  
\affiliation[4]{Faculty of Engineering and Science, Western Norway University of Applied Sciences, Bergen, Norway}   
\affiliation[5]{Institut f\"{u}r Kernphysik, Johann Wolfgang Goethe-Universit\"{a}t Frankfurt, Frankfurt, Germany}   
\affiliation[6]{Lund University Department of Physics, Division of Particle Physics, Lund, Sweden} 
\affiliation[7]{Niels Bohr Institute, University of Copenhagen, Copenhagen, Denmark}   
\affiliation[8]{Physikalisches Institut, Eberhard-Karls-Universit\"{a}t T\"{u}bingen, T\"{u}bingen, Germany}  
\affiliation[9]{Physikalisches Institut, Ruprecht-Karls-Universit\"{a}t Heidelberg, Heidelberg, Germany}  
\affiliation[10]{Research Division and ExtreMe Matter Institute EMMI, GSI Helmholtzzentrum f\"ur Schwerionenforschung GmbH, Darmstadt, Germany}  
\affiliation[11]{The Henryk Niewodniczanski Institute of Nuclear Physics, Polish Academy of Sciences, Cracow, Poland}  
\affiliation[12]{University of Tokyo, Tokyo, Japan}
\affiliation[13]{Yale University, New Haven, Connecticut, United States}
\affiliation[14]{Zentrum  f\"{u}r Technologie und Transfer (ZTT), Worms, Germany}

\author[2]{J.~Alme}
\author[5]{T.~Alt}    
\author[5]{H.~Appelsh\"{a}user}    
\author[13]{M.~Arslandok}   
\author[10]{R.~Averbeck} 
\author[5]{E.~Bartsch}  
\author[10]{P.~Becht}  
\author[5]{L.~Bratrud}    
\author[10]{P.~Braun-Munzinger}   
\author[5]{H.~Buesching}    
\author[13]{H.~Caines} 
\author[6]{P.~Christiansen}  
\author[3]{F.~Costa}    
\author[10]{U.~Frankenfeld}    
\author[7]{J.J.~Gaardh{\o}je}  
\author[10]{C.~Garabatos}  
\author[9]{P.~Gl\"{a}ssel}  
\author[12]{T.~Gunji}
\author[12]{H.~Hamagaki}
\author[13]{J.W.~Harris}  
\author[10]{E.~Hellb\"{a}r}  
\author[4]{H.~Helstrup}         
\author[10]{M.~Ivanov}   
\author[5]{J.~Jung}  
\author[5]{M.~Jung}  
\author[3]{A.~Junique}   
\author[3]{A.~Kalweit} 
\author[14]{R.~Keidel}  
\author[5]{S.~Kirsch}   
\author[5]{M.~Kleiner}  
\author[11]{M.~Kowalski}  
\author[5]{M.~Kr\"uger}  
\author[10]{C.~Lippmann}  
\author[3]{M.~Mager} 
\author[10]{S.~Masciocchi}  
\author[11]{A.~Matyja}  
\author[10]{D.~Mi\'{s}kowiec}   
\author[5]{R.H.~Munzer}  
\author[3]{L.~Musa}  
\author[7]{B.S.~Nielsen}    
\author[11]{J.~Otwinowski}  
\author[1]{M.~Pikna}    
\author[2]{A.~Rehman}         
\author[5]{R.~Renfordt}         
\author[2]{D.~R\"ohrich}       
\author[5]{H.S.~Scheid}  
\author[10]{C.~Schmidt}  
\author[8]{H.R.~Schmidt} 
\author[10]{K.~Schweda}  
\author[12]{ Y.~Sekiguchi}
\author[6]{D.~Silvermyr} 
\author[1]{B.~Sitar}    
\author[9]{J.~Stachel}  
\author[2]{K.~Ullaland}         
\author[3]{R.~Veenhof}  
\author[6]{V.~Vislavicius}  
\author[5]{J.~Wiechula}  
\author[9]{B.~Windelband}